\documentclass[slac_one]{revtex4}
\usepackage{graphicx}
\usepackage{fancyhdr}
\usepackage{wrapfig}
\usepackage{subfigure}
\begin{document}

\title{The Longitudinal Proton Structure Function at HERA} %% Paper title goes here


\author{Jason Schwartz, on behalf of the H1 and ZEUS collaborations}
\affiliation{McGill University, Montreal, Quebec H3A 2T8, Canada}
%\footnote{Presented on behalf of the ZEUS and H1 experiments}
\begin{abstract}
The longitudinal proton structure function ($F_{L}$) has been measured at the HERA collider in positron-proton deep inelastic scattering collisions with the H1 and ZEUS detectors.  This measurement is achieved by using multiple center-of-mass energies ($\sqrt{s}$) via a reduction of the proton beam energy. The energies used for this measurement are $\sqrt{s}$ = 318, 251, 225 GeV.  The kinematic region studied is $2.5<Q^2<800$ $GeV^2$.  H1 and ZEUS data have been combined to increase statistical precision and reduce systematic effects.  The impact of the low proton beam energy cross sections to the proton PDFs is being investigated.
\end{abstract}

\maketitle

\thispagestyle{fancy}


\section{INTRODUCTION} % Section title should be in all capitals.
The $e^{\pm}p$ deep inelastic scattering (DIS) cross section can be represented as the sum of the two structure functions $F_{2}$ and $F_{L}$, as 
\begin{equation} \label{eq:DISXSec}
\sigma_{r} = \frac{d^2\sigma}{dxdQ^2} \frac{Q^4x}{2\pi\alpha^2Y_{+}} = F_{2}(x, Q^2) - \frac{y^2}{Y_{+}}F_{L}(x, Q^{2})
 \end{equation}
where $\alpha$ is the fine structure constant, $Q^2$ is the virtuality of the exchanged boson, $x$ is the Bjorken scaling variable, $y$ is the electron inelasticity and $Y_{+}\equiv 1 + (1-y)^2$.

$F_{2}$ is the dominant term in this cross section throughout most of
the kinematic range.  The magnitude of $F_{L}$ is proportional to the cross
section for probing the proton with a longitudinally polarized virtual
photon ($\sigma_{L}$).

A measurement of $F_{L}$ is technically challenging
and requires a measurement of the DIS cross section at the same $x$
and $Q^2$ while varying $y$.  From the relation $Q^2 = sxy$, it is
clear that this can be achieved only by varying $s$, the center-of-mass energy.
Before HERA ceased operations in July 2007, the proton beam energy was
lowered from its nominal energy of 920\,GeV to 460\,GeV and 575\,GeV to facilitate this measurement.  

\section{THE LONGITUDINAL PROTON STRUCTURE FUNCTION \boldmath{$F_{L}$}}

In the Quark Parton Model (QPM), all of the proton's momentum is carried by the quarks.  $F_{2}$ is equal to the sum of quark and anti-quark $x$ distributions weighted by the square of the electric charges, furthermore $F_{L}$ is exactly zero.  In Quantum Chromodynamics (QCD), $F_{L}$ no longer needs to be zero as it receives contributions from both quarks and gluons.  In DIS at low $x$ the gluon contribution greatly exceeds that of the quarks, making $F_{L}$ a direct measure of the gluon distribution.  Scaling violations in the evolution of $F_{2}$ at low-$x$, as described by the DGLAP QCD evolution equations,  have previously been used to constrain the gluon distribution and $F_{L}$\cite{ref:H1ModDepFL}. An independent measurement of $F_{L}$ will stand to further our knowledge of the longitudinal structure function and act as a validity test of perturbative QCD in the low Bjorken $x$ region.

\section{MEASUREMENT STRATEGY}
\begin{figure}[htbp!]
\begin{center}
\subfigure[ZEUS]{\label{fig:redxsec}\includegraphics[scale=0.35]{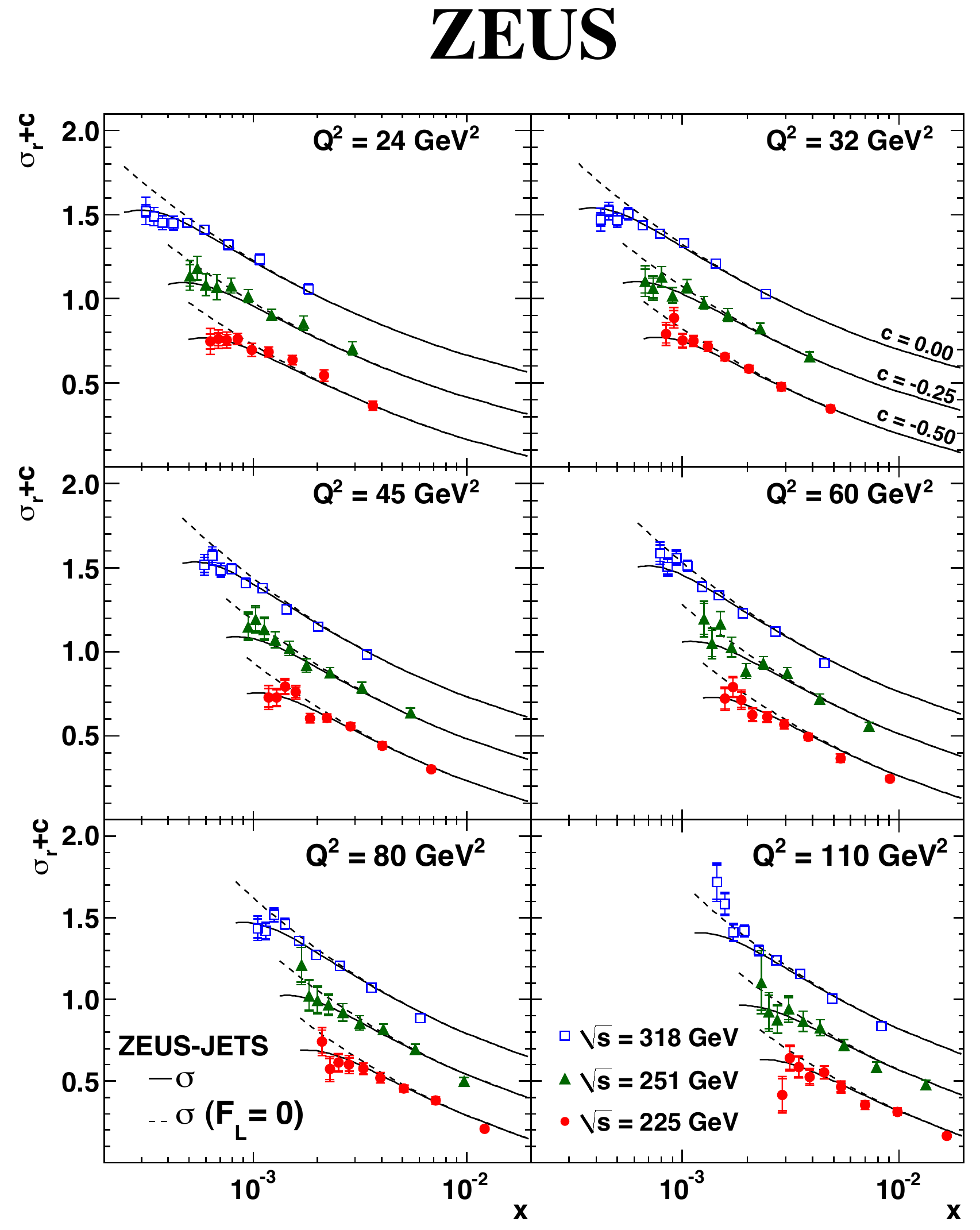}}
\subfigure[H1]{\label{fig:redxsecH1}\includegraphics[scale=0.6]{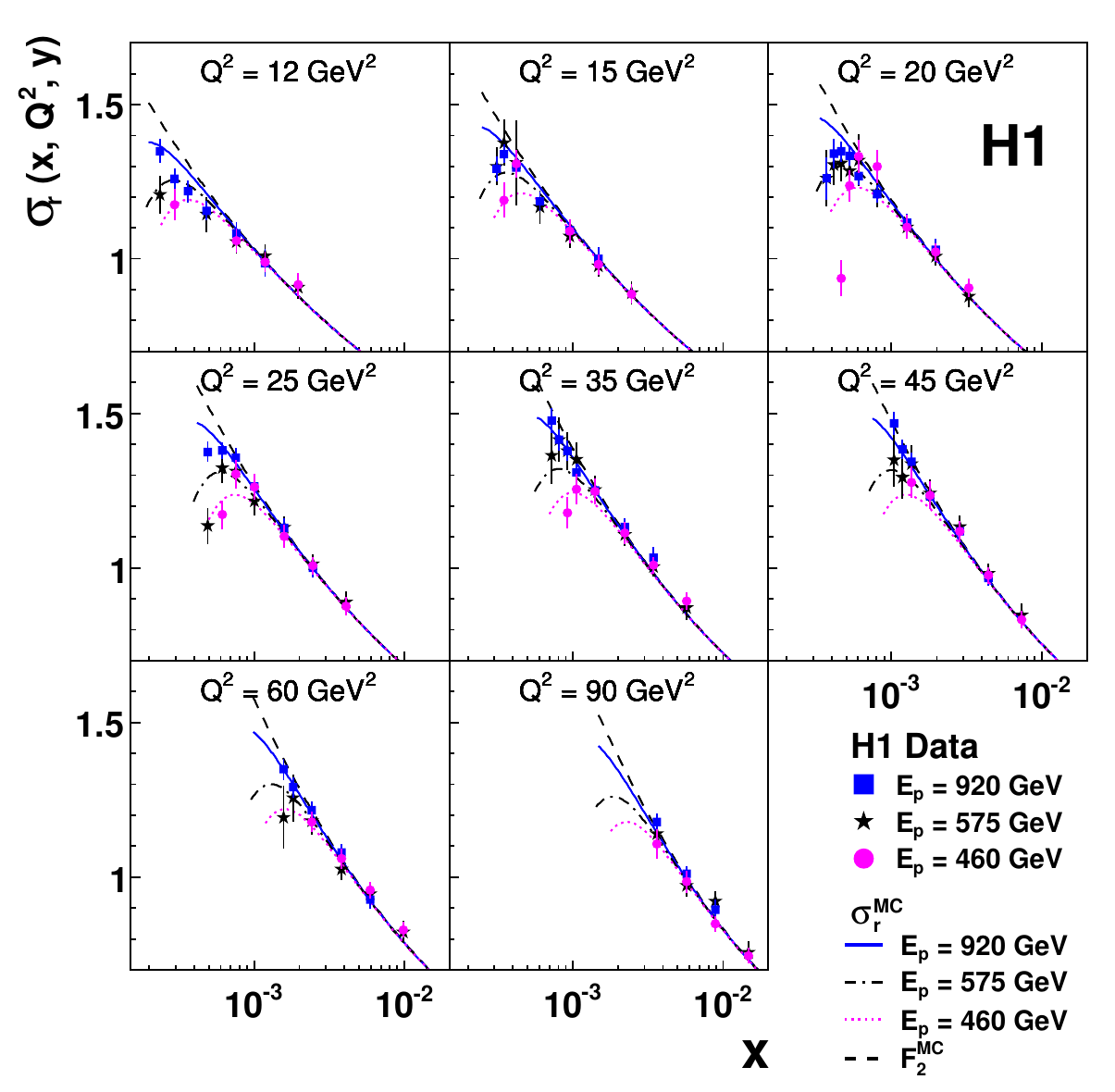}}
\caption{Measured reduced cross section for different bins of $Q^{2}$.}
\end{center}
\end{figure}

\begin{wrapfigure}{r}{0.4\textwidth}
\vspace{-20pt}
  \begin{center}
   \includegraphics[scale=0.60]{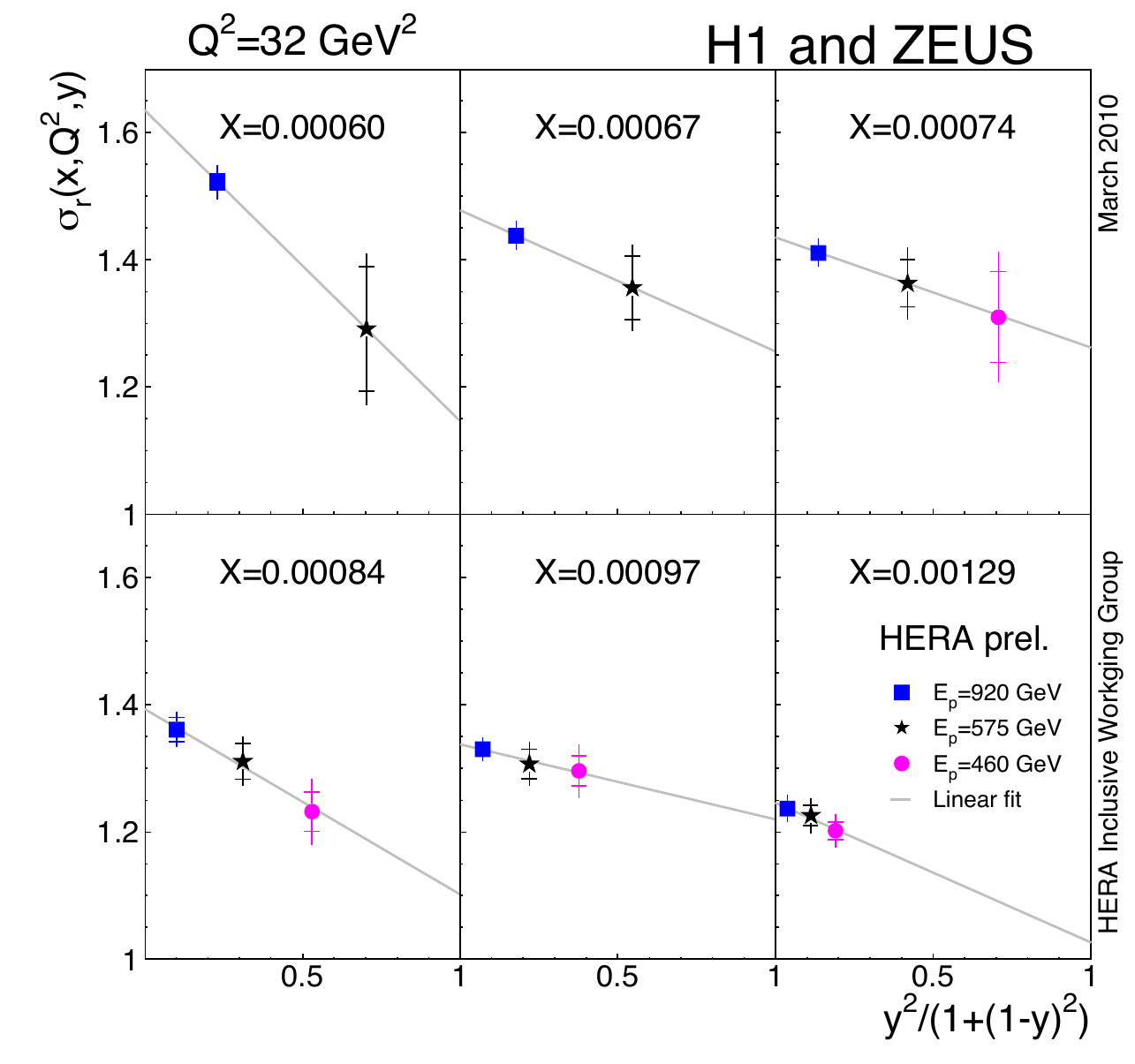}
  \caption{H1 and ZEUS combined measurement of the reduced cross section in a single $Q^{2}$ region for different bins of $x$.  $F_{L}$ is the slope of the line connecting the points.}
  \label{fig:Rosenbluth}
    \end{center}
    \vspace{-40pt}
\end{wrapfigure}

The reduced cross section ($\sigma_{r}$) is measured for multiple beam energies, as is shown in figures \ref{fig:redxsec} and \ref{fig:redxsecH1}.  $F_{L}$ and $F_{2}$ are extracted simultaneously using
a Rosenbluth plot\cite{ref:Rosenbluth}.  The same $\sigma_{r}$ is measured at each centre-of-mass energy and plotted against $y^{2}/Y_{+}$. Figure~\ref{fig:Rosenbluth} demonstrates six such Rosenbluth plots used in the H1 and ZEUS combined measurement.  In a Rosenbluth plot $F_{L}$ is the slope of the line fitted to the points, $F_{L}(x, Q^{2}) = -\partial\sigma_{r}(x, Q^{2}, y)/\partial(y^{2}/Y_{+})$.   $F_{2}$ is simply the $y-$intercept of the said line, $F_{2}(x, Q^{2}) = \sigma_{r}(x, Q^{2}, y = 0)$.

%\subfigure[]{\label{fig:Rosenbluth}\includegraphics[scale=0.50]{FLFit.pdf}}

Both the H1 and ZEUS collaborations have independently measured $F_{L}$\cite{ref:ZEUSFL, ref:H1FL}.  Combining the measurements from the two experiments has several advantages, firstly, an increase in statistics improves the
precision of the measurement. Secondly, uncorrelated errors reduce because they are different for both H1 and ZEUS.  The point-to-point correlated errors also reduce because H1 and ZEUS use different methods for reconstructing the event kinematics.  To combine the data, results from H1 and ZEUS are corrected to a common $x$, $Q^{2}$ grid.  This is done using a swimming technique involving the HERAPDF 1.0 parameterization \cite{ref:HERAPDF}.  The reduced cross sections are combined and then $F_{L}$ is extracted. 

\begin{figure}
\vspace{-10pt}
  \begin{center}
   \includegraphics[scale=0.70]{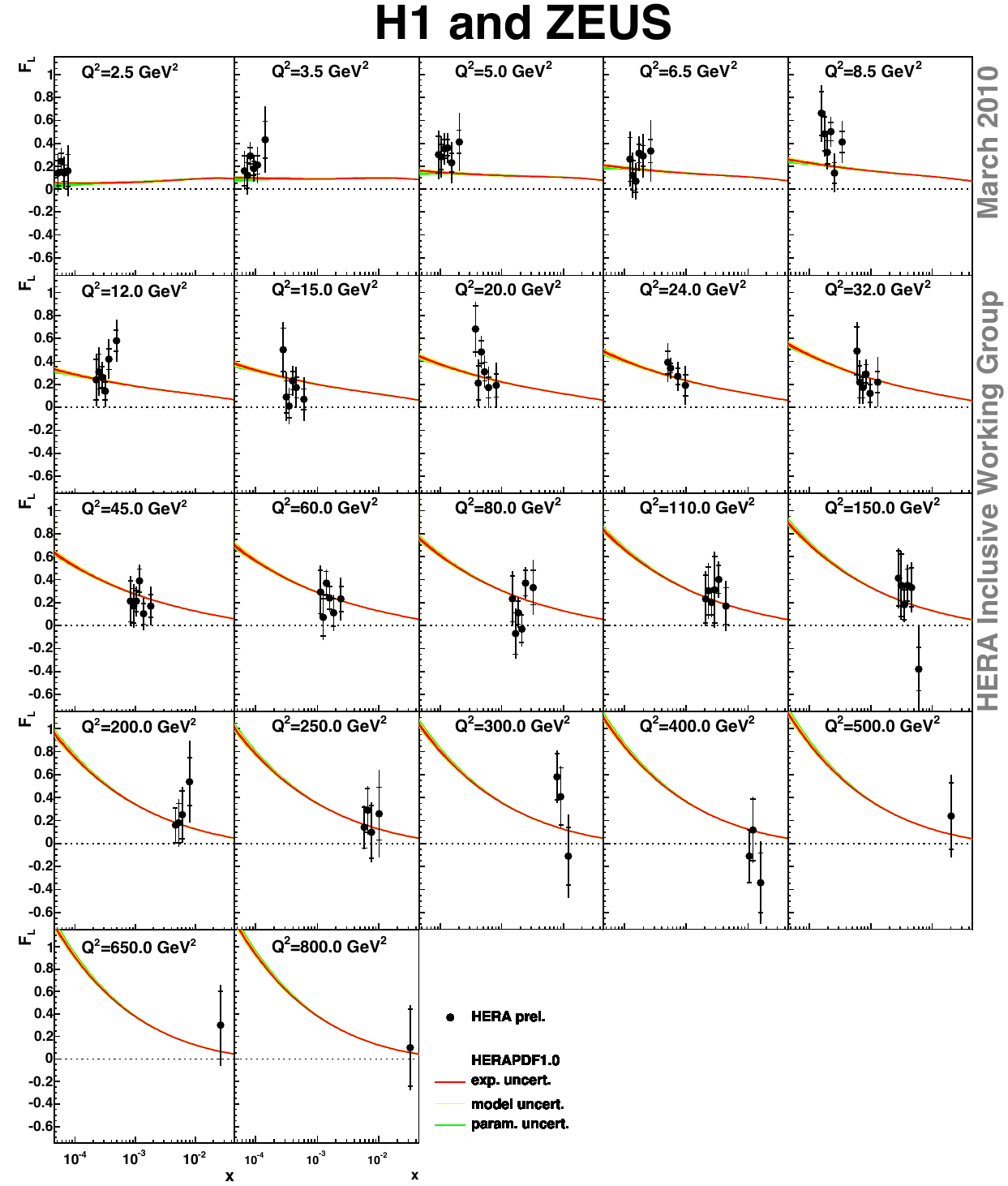}
  \caption{Combined H1 and ZEUS $F_{L}$ measurement.}
  \label{fig:FL}
    \end{center}
    \vspace{-10pt}
\end{figure}
\section{RESULTS}
\begin{figure}[htbp!]
    \vspace{-10pt}
\begin{center}
\subfigure[]{\label{fig:H1ZEUS_Q2}\includegraphics[scale=0.80]{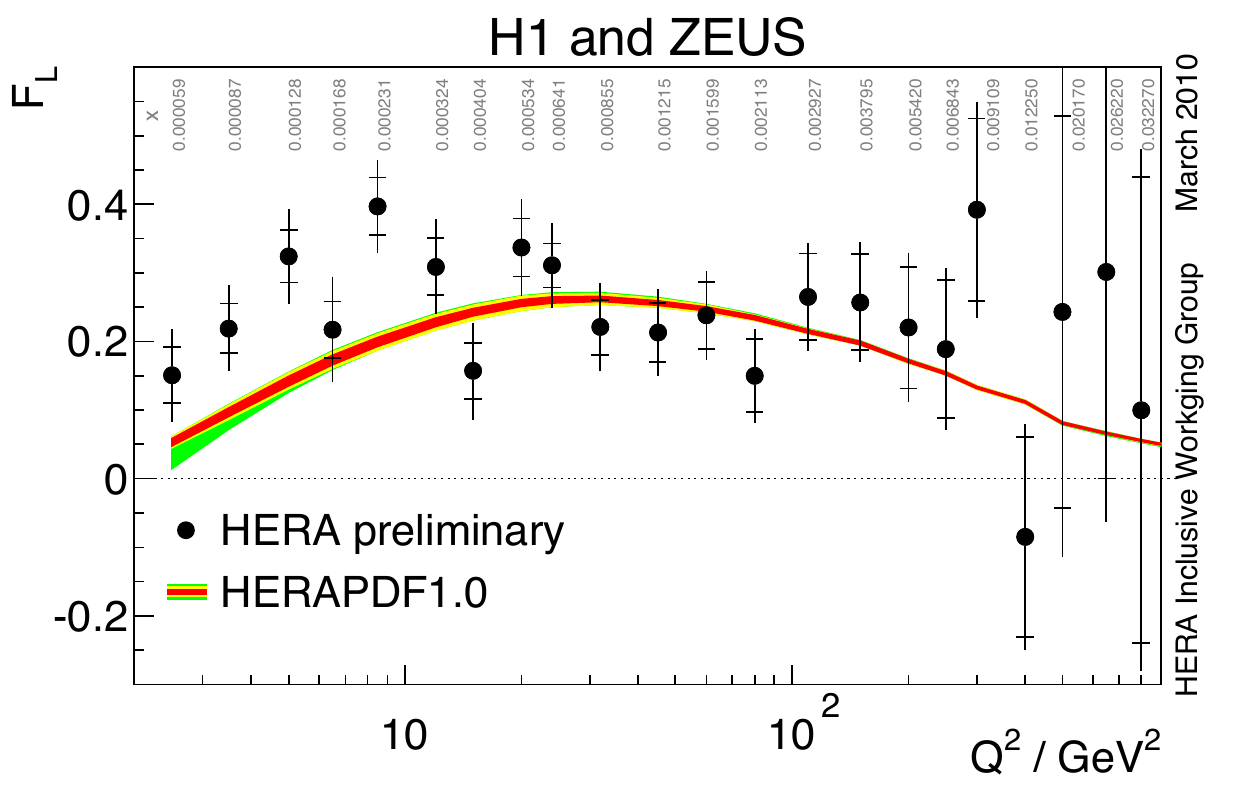}}
\subfigure[]{\label{fig:PartonPDF}\includegraphics[scale=0.35]{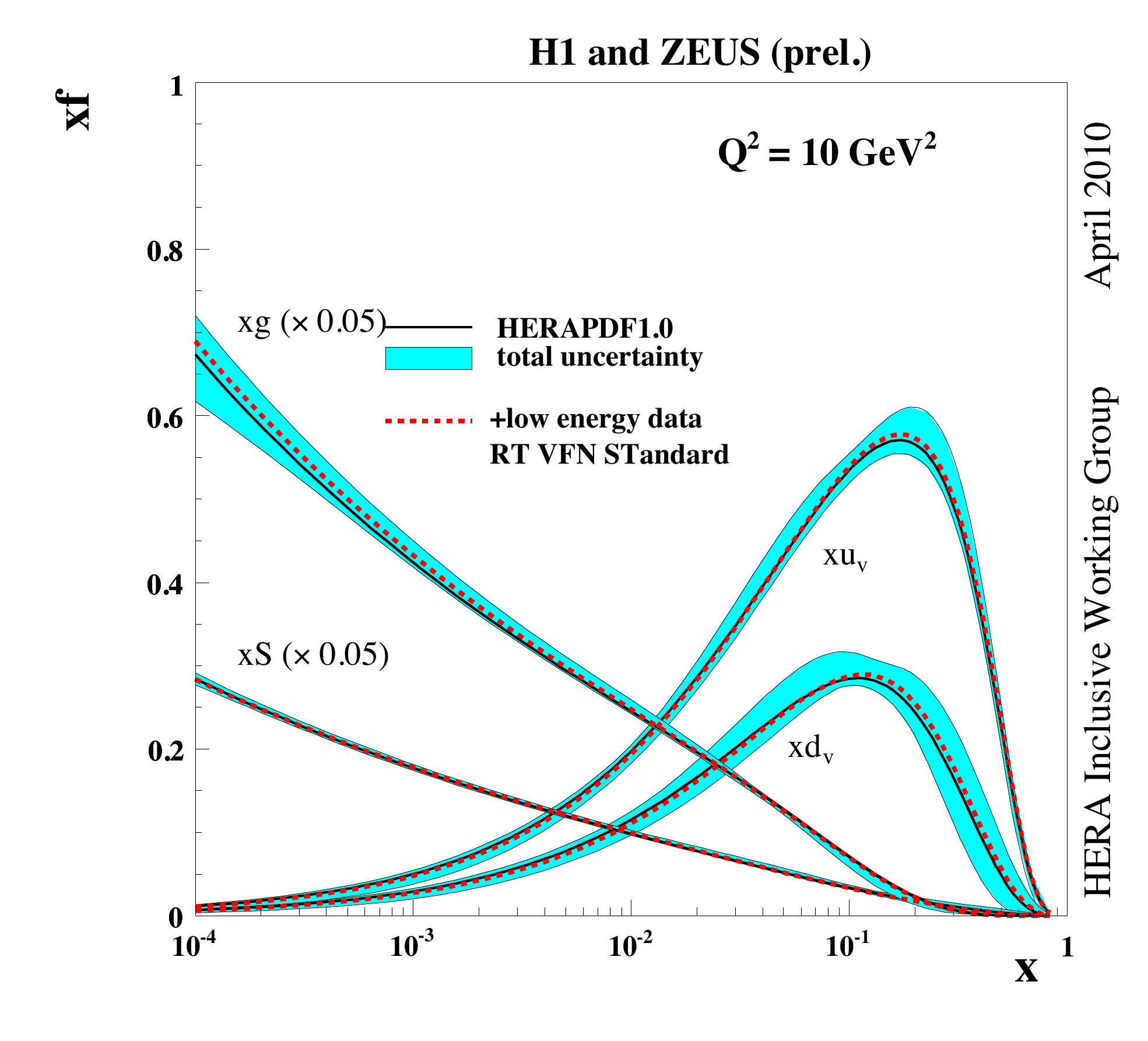}}
\caption{\ref{fig:H1ZEUS_Q2} Combined H1 and ZEUS $F_{L}$ measurement as a function of $Q^{2}$ averaged over $x$. \ref{fig:PartonPDF} Proton Parton Distribution Functions (PDF) for u, d, Sea and Gluons.	Dashed line represents PDFs with low energy cross sections included.}
\end{center}
 \vspace{-15pt}
\end{figure}

This measurement covers a wide kinematic range, spanning $2.5 < Q^{2} < 800 \text{GeV}^{2}$ and $0.0006 < x < 0.0036$.  The combined results for $F_{L}$ are shown in figure~\ref{fig:FL}.  In figure~\ref{fig:H1ZEUS_Q2} each $Q^{2}$ bin is averaged over all the measured $x$ values to obtain a single $x$ bin.  The results are compared with the HERAPDF 1.0 PDF set. Good agreement is demonstrated for $Q^{2} > 10\text{GeV}^2$ while at low $Q^{2}$ measurements deviate from NLO QCD predictions.  In general results are consistent with non-zero $F_{L}$.  H1 and ZEUS combined results are now being used to improve the HERAPDF 1.0 parameterization sets.  Figure~\ref{fig:PartonPDF} is an example of one such parameterization and compares the PDF set with and without the inclusion of $\sqrt{s}$ = 251 \& 225 GeV data.  The two PDF sets agree within the total uncertainty of HERAPDF 1.0, supporting the expectations of perturbative QCD. 

%\begin{acknowledgments}
%The authors wish to thank JACoW for their guidance in preparing this template.

%Work supported by Department of Energy contract DE-AC02-76SF00515.
%\end{acknowledgments}

\end{document}